# Bayesian Analogical Cybernetics

Adam Safron

Indiana University

**Abstract**

It has been argued that all of cognition can be understood in terms of Bayesian inference. It has also been argued that analogy is the core of cognition. Here I will propose that these perspectives are fully compatible, in that analogical reasoning can be described in terms of Bayesian inference and vice versa, and that both of these positions require a thorough cybernetic grounding in order to fulfill their promise as unifying frameworks for understanding minds. From the Bayesian perspective of the Free Energy Principle and Active Inference framework, thought is constituted by dynamics of cascading belief propagation through the nodes of probabilistic generative models specified by a cortical heterarchy "rooted" in action-perception cycles that ground the mind as an embodied control system for an autonomous agent. From the analogical structure mapping perspective, thought is constituted by the alignment and comparison of heterogeneous structural representations. Here I will propose that this core cognitive process for analogical reasoning is naturally implemented by predictive coding mechanisms. However, both Bayesian cognitive science and models of cognitive development via analogical reasoning require rich base domains and priors (or reliably learnable posteriors) from which they can commence the process of bootstrapping minds. Here in the spirit of the work of George Lakoff and Mark Johnson, I propose that embodiment provides many of the inductive biases that are usually described in terms of innate core knowledge. *(Please note: this manuscript was written and finalized in 2012)*

**Table of Contents**

## Philosophical background and goals

*"[Science can be likened] to a boat which, if we are to rebuild it, we must rebuild plank by plank while staying afloat in it. The philosopher and the scientist are in the same boat. Our boat stays afloat because at each alteration we keep the bulk of it intact as a going concern."*

–*Willard van Orman Quine (1908-2000), discussing Otto Neurath (1882-1945)*

In attempting to understand the mind, the erroneous philosophical positions of dualism and eliminative reductionism are avoided by recognizing that cognition can be explained in terms of multiple compatible levels of analysis (D. C. Dennett, 2003; Hofstadter, 1979). Fundamentally, an aspect of the mind is not understood unless it can be described from multiple perspectives, as well as their particular correspondences. This sort of epistemological pluralism allows us to more fully specify the meaning of our constructs, and thus minimize conceptual ambiguity.

David Marr famously proposed three complementary levels of analysis for explaining cognitive phenomena: 1) the computational level describes what a system does and why it does it; 2) the algorithmic level describes how a system accomplishes these functions through processes that build and manipulate representations; 3) the implementation level describes how these processes and representations are physically realized, whether by nervous systems or computer-based intelligences (J. P. Mitchell, 2006). Although this paper will primarily use Marr's language, it is important to note that an unbounded number of valid perspectives are possible (Ziporyn, 2004), which are more or less effective at "carving nature at its joints" such that relevant emergent phenomena can be comprehended.[1] For instance, an alternative set of joints can be found if one considers behavior, cognitive events, neural events, and experience as different aspects of multi-faceted minds.

However, these different views are incommensurate without using probability theory and the language of computation as translating principles. Here, we describe a synthetic approach to understanding neuropsychological phenomena in terms of a perspective that integrates dynamical systems theory and machine learning within a "Bayesian-Analogical Cybernetics" (BAC) framework. BAC describes all cognitive processes in terms of Bayesian inference, analogical reasoning, and the cybernetic perspective of minds as control systems for embodied agents, which develop through their interactions with environments.

---

[1] In *The Phaedrus,* the Platonic dialogue from which the expression of "carving nature at its joints" is derived (Griswold, 1996), Socrates likens the philosopher to a butcher seeking to divide an animal in ways that are anatomically sensible. But even here, the particular joints identified are partially dependent on pragmatic factors, or simply the limited-pragmatism of tradition, as can be verified by seeing the diversity of cuts of meat produced by butchers of different cultures. However, variation of classification schemes does not mean that all conceptualizations are similarly valid in all contexts.

# Bayesian cognitive science

## A whirlwind tour of Bayesian inference; from basic hypothesis testing to computer simulations

*"Probability theory is nothing but common sense reduced to calculation."*

> *–Pierre-Simon Laplace (1749–1827)*

Simply, we will use the term "Bayesian" to describe all methods that attempt to combine and update conditional probabilities on the basis of prior beliefs and evidence. Less simply, this section will provide an introduction to Bayesian methods of probabilistic inference for which there are five important points that it would be helpful for the reader to understand before proceeding to the next section:

1. Bayesian inference is based on conditional probabilities: i.e., degrees of certainty for which the order of specification matters; e.g. "what is the probability that the ground is wet, given that there are clouds" vs. "what is the probability that it is raining, given that the ground is wet."
2. Bayesian methods can be used to combine beliefs from past experiences (i.e., prior probabilities) with current observations, and then update these prior beliefs to obtain empirically tested revised beliefs (i.e., posterior probabilities). These updated prior beliefs—now known as the posterior probability distribution—provide new priors for further hypothesis refinement.
3. Bayesian model selection involves a process where different hypotheses, and systems of hypotheses, compete to explain the greatest amount of observations with the minimum amount of model complexity (i.e., parsimony).
4. Elaborate hypothesis systems can be represented using graphical models, whose potentially complex structures can specify intricate dependencies among conditional probabilities.
5. In order to generate inferences from graphical models, computer simulation techniques such as Markov chain Monte Carlo (MCMC) use semi-random sampling to estimate associated variable-specific (i.e., "marginal") and combined (i.e., "joint") probability distributions.

"Bayesian" approaches are distinguished by using prior knowledge to help interpret observed data, and then updating this knowledge on the basis of observations (Cronin, Stevenson, Sur, & Körding, 2010): Which hypotheses could plausibly explain different aspects of the data? Before making observations, a priori, what is the probability that these hypotheses will be supported? What is the likelihood for a hypothesis, given the data? What is the relative likelihood for a hypothesis, given all of the potential alternative hypotheses and their associated prior probabilities and likelihoods? What causal relations might influence the prior probabilities and likelihoods associated with these hypotheses?

By asking these questions, investigators can more effectively select models and estimate associated parameter values and confidence intervals, given a particular set of observations, and given the set of alternative model candidates. More specifically, Bayesian model selection starts by explicitly postulating dependencies between variables of interest, as well as uncertainty associated with these relations. In considering multiple hypotheses simultaneously in the joint posterior distribution—integrated using Bayes' rule—not only do these methods take into account more relevant information during inference, but they also provide detailed descriptions of the range of alternative explanations.

The relative plausibility of hypotheses can be assessed through a variety of methods, such as using the ratio of marginal probabilities associated with each model to calculate a "Bayes Factor." The rules of probabilistic inference ensure that additional degrees of freedom are penalized during model selection, so instantiating "Occam's Razor," or the principle of parsimony. Bayesian methods most often produce similar results to more frequently used techniques, but informative prior probabilities and valid models allow inferences to be made from less data with greater certainty. Although inaccurate priors can interfere with valid inference, this problem can potentially be remedied with sufficient time and observations, as beliefs are updated on the basis of further evidence.

Bayesian networks allow statistical inference to be performed with complex systems of conditional probabilities, which can be designed to reflect the structured relations of empirical phenomena. These models are "generative" in that they specify posterior marginal and joint probability distributions (i.e., probability density maps) resulting from the combination of all available relevant information. By using the structure of relationships among conditional probabilities, probabilistic graphical models increase inferential power by reducing the number of potential models that need to be considered.

Before the advent of digital computers, Bayesian modeling was limited due to the infeasibility of performing the required computations. For generative models of any significant degree of complexity, the associated integrals are frequently intractable using analytic methods. Additionally, probability density maps can be impossible to visualize when they result in topological manifolds in multi-dimensional spaces. Fortunately, computer-aided sampling techniques can reliably approximate solutions by performing biased searches of estimated posterior probability distributions.

With Markov chain Monte Carlo (MCMC) simulations, Markov chains—characterized by a "memoryless" "random walk" in that the algorithm is not affected by previous states—explore high-dimensional spaces by proposing randomly altered parameter values. These potential "steps" are evaluated using Bayes rule to calculate associated probabilities. If a proposal would result in a higher probability relative to the current state, the algorithm changes its current estimates and the chain moves to that region of parameter space. If the move would result in a lower probability relative to the current state, the algorithm either maintains its current position, or possibly takes a step based on probabilistic decision criteria.

After sufficient iterations, time spent at different areas of the search space becomes proportional to the previously non-calculable, model-specified joint probability density for different permutations of parameter values. MCMC can estimate probability distributions for complex models with arbitrary precision, given enough sampling time.[2] Even more, it can be used to estimate specific parameter values, and the rank ordering of sampled states can be used to create credibility intervals for different tests, even with non-Gaussian posterior distributions.

In addition to MCMC, analytic approaches based on variational (or approximate) inference are increasingly being used for their computational efficiency. Although these will not be discussed here, interested readers would do well to explore the rich literature that has accumulated on these methods and their applications.

### A whirlwind tour of Bayesian cognitive science; from functionalist characterization of complex probabilistic inference, to dynamic systems of evolving hypotheses

"*The probability of any event is the ratio between the value at which an expectation depending on the happening of the event ought to be computed, and the value of the thing expected upon its happening.*"

　　–*Thomas Bayes (1701-1761)*

Simply, Bayesian cognitive science is an approach to understanding mental processes in terms of probabilistic inference. Less simply, this section will provide an overview of recent advances, for which there are five important points that it would be helpful for the reader to understand before proceeding to the next section:

1. Bayesian cognitive science is a diverse field that focuses on using probabilistic inference to characterize problems solved by biological learners on Marr's computational level of analysis. More recently, Bayesian cognitive scientists have also begun to use methods from statistical machine learning to describe psychological phenomena in algorithmic terms.
2. "Infinite" (i.e., nonparametric) methods allow the structure of a variety of complex models to be discovered and modified on the basis of experience, as long as minimal initial knowledge is provided.
3. The origins of this initial knowledge in biological learners are currently poorly understood, but the central role of embodiment will be described in the discussions that follow this section.
4. The field is limited by a lack of integration between computational/algorithmic models and details of neural implementation.
5. When combined with a cybernetics perspective, recent insights from computational neuroscience may provide a means of solving the challenge of integrating Bayesian models with biological mechanisms.

For point #4, we recommend reading the article "Bayesian Fundamentalism or Enlightenment? On the explanatory status and theoretical contributions of Bayesian models of cognition" (M. Jones & Love, 2011). For point #5, we recommend continuing to read this paper.

---

[2] However, with more complex models, in order to estimate representative solutions within polynomial time, it becomes important to design algorithms where proposal distributions efficiently converge upon anticipated sampling distributions.

From a "Bayesian" perspective, all aspects of thought can be understood as probabilistic inferences about "candidate world structures" (Tenenbaum, Kemp, Griffiths, & Goodman, 2011). From this perspective, the mind is a collection of evolving probabilistic generative models that attempt to discover and characterize aspects of the world that cannot be directly observed (i.e., "latent" or "hidden" variables). More specifically, in developing these generative models through repeated observations, model-structure implies generalized knowledge, which specifies the broader class of situations to which future learning should generalize.

Bayesian models are extremely flexible in their representational abilities. As previously described, graph structures (e.g. trees) provide constraints on model development, which increases efficiency by restricting the number of hypotheses that need to be considered. The logic of the graph can eliminate some classes of hypotheses a priori, thus setting their prior probabilities to 0. With smaller hypothesis spaces provided by the constraining structure of graphs, learners are able to generalize far more accurately, assuming that the graphical model is correct. To the extent that aspects of a graphical model are inaccurate, however, they reduce the learner's ability to make valid inferences, but these models can be updated on the basis of further observations. With recent advances, computer programs can be used to produce complex models on the basis of graph grammars, or even statements in first order logic.

By embedding these graphs in hierarchical networks, it is possible to discover the correct structural forms (i.e., grammars) as well as the particular structures (i.e., graphs) that describe conditional probabilistic associations between variables that represent the higher-order relations of diverse "real world" phenomena:

> "*Hierarchical Bayesian models* (*HBM*s) address the origins of hypothesis spaces and priors by positing not just a single level of hypotheses to explain the data but multiple levels: hypothesis spaces of hypothesis spaces, with priors on priors. Each level of a HBM generates a probability distribution on variables at the level below. Bayesian inference across all levels allows hypotheses and priors needed for a specific learning task to themselves be learned at larger or longer time scales, at the same time as they constrain lower-level learning" (Tenenbaum et al., 2011).

Not only can these methods be used to facilitate particular inductive inferences, but they can also discover the most appropriate graph structures for representing the relations between objects and events. Additionally, structures are considered to be psychologically plausible both in terms of the face validity of graph representations, as well as the fact that their relative estimates of item similarity are comparable with the judgments of human subjects.

With "infinite" or nonparametric hierarchical modeling, not even the number of variable classes needs to be known in advance. These methods balance inductive constraints and flexibility by positing an unbounded amount of structure, restricting the degrees of freedom per dataset, and utilizing Bayesian model-selection to automatically apply Occam's razor in network development. Nonparametric models can learn quickly from little data compared to what would be needed at lower levels of pattern abstraction. This feat is achieved by using degrees of freedom at higher levels to pool evidence from many hierarchically lower variables, a feature that Tenenbaum et al. (2011) refer to as the "blessing of abstraction."

Bayesian cognitive scientists believe that these methods of hierarchical model development are the most promising approach to understanding how humans learn and use abstract knowledge. However, these methods may both underestimate and overestimate the potential of probabilistic inference based on graphical models. Regarding overestimation, without the inductive constraints offered by embodiment—i.e., providing a toehold for the bootstrapping process that overcomes the challenge of impoverished stimuli—machine learning approaches will fail to develop sufficiently robust causal reasoning for broad applicability (Clark, 2008; Hirose, 2002; Pezzulo et al., 2011). Regarding underestimation, biologically-inspired algorithms—where cortical columns represent nodes in a particular kind of directed graphical model—may provide a common computational platform for understanding human cognition, as well as for developing flexible and general artificial intelligences (Friston, 2008; Hawkins, 2011; Hawkins & Blakeslee, 2004).

## Analogy as the core of cognition

*"All models are wrong, but some are useful."*

> *–George E. P. Box (1919 - present)*

The promise of nonparametric Bayesian methods is strongly evidenced by their ability to develop complex models from limited observations, which support novel inferences on the basis of generalizable, structured knowledge representations. Regardless of whether or not conditional probabilities are emphasized, these capacities are generally recognized as fundamental aspects of intelligence, and the ability of analogical reasoning to support these processes has lead many cognitive scientists to suggest that analogy is the core of cognition (Flusberg, Thibodeau, Sternberg, & Glick, 2010; Gentner, 2010; Hummel & Holyoak, 1997; Lakoff & Johnson, 1999; Larkey & Love, 2003; M. Mitchell & Hofstadter, 1990). In the following section, we will describe how models of analogical inference achieve computational level goals shared with Bayesian approaches. Then, we will discuss how common details of biological implementation may explain both probabilistic inference and analogical reasoning.

### What is analogy?

Within the BAC framework, *analogy* is defined as any process that generates *inferences* (i.e., probabilistic implications) based on *representations*, which are broadly construed as *structures* whose features *correspond* to aspects of *referenced* phenomena. These referents can be aspects of the world (i.e., latent variables), implicit or explicit models of those aspects (i.e., estimated variables and parameters), or even models of models. In addition to implication driven by representational-structure, we characterize analogies in terms of two inter-related qualitative dimensions of *similarity* and *metaphoricity*. We define *similarity* as the extent to which structures have shared features—compared with unshared features—within a representational system. Thus similarity evaluations are relative to the similarities of all other potentially relevant analogies. Further, we define *metaphoricity* as the extent to which representations must be *restructured* in order to produce analogies of a given degree of metaphoricity. Thus metaphoricity evaluations are functions of particular ranges of similarity values, the relevance of which are determined by contextual factors ranging from system processing limitations, to high-level pragmatic considerations.

In addition to these qualitative dimensions of similarity and metaphoricity, a quantitative dimension of *certainty* can be estimated for particular inferences, as well as systems of inferences. From this perspective, all models are analogies with varying degrees of relative similarity and metaphoricity, as well as inferential certainty. This applies to both individual beliefs as well as scientific hypotheses: representations are never identical with the phenomena to which they refer; also, aspects of the world cannot be known directly, but must be inferred on the basis of limited observations. In this way, all cognition is inherently probabilistic, and so analogies, representations, and models are fundamentally similar concepts. Or, as Korzybski noted, "A map is not the territory… A map covers not all of the territory… A map is self-reflexive" (Blake & Ramsey, 1951). In other words, maps are analogies for the territories they represent, and models are analogies for the phenomena they describe.

Thus, for any given domain, there are countless models with varying degrees of validity and reliability. In the language of analogy, the quality of a model corresponds to the degree of similarity between representations specified by hypotheses, compared with those generated by observations. Further, hypothesis testing uses the structure of a priori models to predict the structure of representations that are likely to result from experiments. The only difference between Bayesian and analogical approaches to inference is whether or not probabilities are explicitly calculated. Indeed, not only do Bayesian models consist of systems of analogies, but analogical reasoning can also be mathematically formalized in terms of Bayesian inference. Thus, there are multiple ways in which Bayesian cognitive science supports the claim that analogy is the core of cognition.[3]

## Models of analogy

Numerous computational models of analogical reasoning have been developed (Gentner & Forbus, 2011; Kokinov & French, 2006; Lakoff, 2009; C. T. Morrison & Dietrich, 1995; Murdock, 2011), each with specific advantages and disadvantages. *Symbolic* approaches are particularly effective for problems involving high-level cognitive processing. *Connectionist* approaches, on the other hand, attempt to use excitatory and inhibitory associations between nodes in artificial neural networks for discovering patterns, and then manipulating those patterns such that cognitive-level goals can be realized by sub-symbolic mechanisms that are "biologically plausible." Although these quasi-biological approaches have mild success when it comes to processing unstructured inputs, they tend to be limited in handling the high-level cognitive processes that are the strength of symbolic models. Alternatively, *hybrid* models of analogical reasoning combine symbolic and connectionist features in an attempt avoid their individual weaknesses while keeping their relative strengths. The BAC framework supports a hybrid model where the symbolic level emerges from a biologically realistic "connectionist" architecture that initially lacks symbolic representations.

---

[3] Analogical thinking is often considered to be a relatively unreliable mode of reasoning, but this belief both underestimates analogy and overestimates reason. While we may have absolute confidence in the rules of logic, complete certainty is lost as soon as we attempt to ground meanings with referents. Although this probabilistic epistemology eliminates the possibility of certain knowledge, in exchange, it provides a foundation for developing increasingly justified beliefs through experience.

These approaches vary in important ways, but their similarities provide convergent validity for analogy being a fundamental property of minds. Next, we will describe a particularly influential symbolic model of analogical reasoning: *Structure Mapping Theory* (SMT) (Gentner, 1983). Although other models have attempted to incorporate biological constraints (Hummel & Holyoak, 1997; Larkey & Love, 2003; M. Mitchell & Hofstadter, 1990; Petrov, 2000), SMT is particularly well-suited for describing a uniquely human form of analogical reasoning that emphasizes high-level relations among explicitly represented symbols (Premack, 1983).

## Structure Mapping Theory

SMT was the first model of analogical reasoning to identify the importance of structural similarity between representations, as determined by common systems of relations across domains (Gentner, 1983). The theory suggests that humans make comparisons by establishing structural alignments between pairs of represented situations and projecting inferences between a source/base domain that is more detailed and a target domain that is less detailed. These representations consist of connected systems of different kinds of elements and their attributes, including specifications of objects, relations among these objects, and relations of relations. A maximally consistent alignment between representations identifies commonalities and differences, which are used to estimate the degree of similarity between mapped structures.

On the basis of extensive research in humans, Gentner identified several essential phenomena for characterizing analogical processing (Gentner, 2006):

1. *Relational similarity*: relational commonalities are more important than object commonalities in evaluating the similarity of representational structures.
2. *Structural consistency*: representations are aligned by establishing *one-to-one mappings* of structure elements, while maintaining *parallel connectivity* of relations among mapped elements in each system.
3. *Systematicity*: mappings are most significant when they align connected systems of relations, with more connected mappings having greater influences.
4. *Candidate inferences*: if a base domain suggests a missing relation in the target domain, these relations can be inferred using "pattern completion."
5. *Alignable differences*: mapped differences are more salient if connected to shared systems of relations, and thus impact overall systematicity.

6. *Interactive mapping*: the mapping process is determined based on the particular domains being compared, thus different pairings can result in different interpretations.
7. *Multiple interpretations*: a single comparison can lead to more than one interpretation for a given structural alignment.
8. *Cross-mapping*: when object features suggest opposing interpretations to relational features, analogies are more difficult to process.

Any adequate account of analogical reasoning must be able to explain these empirically verifiable phenomena, as well as other significant factors (e.g. emotional salience). Although a thorough analysis is beyond the scope of this paper, nearly every detail of SMT-based algorithms and cognitive models can also be described in terms of statistical machine learning and computational neuroscience.[4] In light of the fact that SMT was developed without considering details of biological implementation or probability theory, this compatibility provides convergent support for the BAC framework.

The *Structure Mapping Engine* (SME) is a computational model that generates analogies according to the principles of SMT (Forbus et al., 1986). SME aligns representations by finding all local identity matches between the two systems in parallel, and then combining them into structurally consistent candidate clusters (i.e., kernels). These clusters of structures are then merged in a process that favors the largest kernels with the most systematicity, determined by the degree to which higher-order relations connect lower-order matches. Aligned representations with greater systematicity maximize overall structural evaluation scores, which indicate preferred mappings.

More specifically, SME begins by attempting to construct domain representations in a bottom-up fashion through a parallel search where all possible matches between domains are considered. These initial representations must satisfy the constraint of "*tiered identicality*" in that element pairings are only considered valid if the elements consist of identical relations. These representation sub-structures are then evaluated according to the parallel connectivity constraint, which requires that alignments between domains maintain consistent patterns of connectivity. The result of this process is a "*hypothesis forest*" that is grouped into maximally consistent clusters of match hypotheses, which are then sequentially combined to create final mappings. Hypotheses are combined using a "*greedy merge*" algorithm that selects the largest kernel and adds as many other kernels as possible. The systematicity bias favors mappings with higher-level relations that support "deep" explanatory or causal structure. Finally, if elements are present in the more complete base domain but absent in the less complete domain, then pattern completion fills in candidate inferences from the base to the target.

In attempting to create alignments with maximum structural consistency, sometimes domains need to be restructured in order to discover relevant commonalities that may be missed based on initial representations. In the previously described process of forming representations, specific elements are counted as "matched" only if they satisfy the constraint that relationships between elements are identical. If a highly structurally consistent alignment could be obtained except for a single violation of this "tiered identicality" constraint, then under some circumstances an attempt is made to force a match. With the technique of "*minimum ascension,*" unmatched features are replaced with the categories from the next higher level of a taxonomic hierarchy of representational elements.

---

[4] For example, the analogical systematicity bias can be explained in terms of Bayesian model selection, or it could also be explained in terms of coherent information processing being selected by a common cortical algorithm that minimizes prediction-errors.

Gentner and colleagues explain this process of loosening criteria as constituting a tradeoff where chance false positives are allowed for the sake of avoiding the false negatives of missed potential matches (Yan, 2003). Flexible analogical mapping enhances both the creativity and robustness of reasoning:[5]

> "Whatever internal representation is generated for seeing, for example, a cat is expressed in the same internal representational conventions from one instant to the next, although the details of the specific descriptions computed may change as the cat stretches…. The specific contents of descriptions for two distinct cats, for example, might vary widely due to differences in what was attended to as well as differences between the cats themselves, but it seems likely that much of the basic vocabulary of perceptual and physical relationships is roughly constant over time. On the other hand, differences in attention and task demands will affect what is encoded and to some degree how, and learning can change conceptual vocabularies and encoding strategies."

However, SME prioritizes false positive avoidance over missed mapping opportunities, particularly for concrete descriptions. False matches require considering more correspondences in the process of merging local hypotheses to form analogical mappings, which could strain processing resources. Rather than allowing for overly flexible initial mappings, SME examines pre-existing mappings to see if results are acceptable by pragmatic considerations relevant to the current task. If initially produced mappings are found to be suboptimal, the process of "rerepresentation" involves performing minor alterations of the base and target domains to see if an improved alignment is achievable.

### Structure mapping and the mind

*MAC/FAC* (Forbus, Gentner, & Law, 1994) is a SME-based model of similarity-guided retrieval that accounts for ways that human analogical reasoning deviates from theoretical optimality. In accordance with SMT, individuals emphasize structural commonalities for similarity judgments on items held in working memory. However these same participants focus on superficial commonalities when retrieving similar examples from long-term memory, even though structure tends to be favored in subjective evaluations. MAC/FAC attempts to explain this discrepancy by separating memory-based analogical mapping into a retrieval stage and a selection stage. In the MAC (*Many are Called*) stage, potential base systems are retrieved from a long-term memory as a function of the magnitude of dot products between the content vector descriptions of the systems in the long-term memory knowledgebase and the contents of working memory (i.e., the probes). This method of retrieval produces candidate base domains with high surface similarity relative to probes. In the FAC (*Few are Chosen*) phase, SME creates and evaluates mappings between base and target systems and chooses solutions with the greatest overall similarity. If FAC chooses between potential base systems that are dominated by surface similarity in the MAC phase, then selected mappings will be similarly limited. Additionally, any factor that impairs the selection phase—whether by compromising working memory or restricting the amount of time available for evaluating mappings—will also compromise the quality of analogies and their associated inferences.

---

[5] When applied to category formation as in the model described below, this flexibility may aid in the formation of invariant representations.

*SAGE (the program formerly known as SEQL)* is a SME-based model of category learning that forms abstractions by making successive structural comparisons among exemplars (Kuehne, Forbus, Gentner, & Quinn, 2000). New exemplars are first compared to existing abstractions, and tested for whether the degree of similarity—determined by structural evaluation scores normalized by base-description size—is greater than a modeler-specified threshold for combination. If this similarity-based combination threshold is exceeded, the abstraction is modified to represent the intersection formed via alignment, and becomes increasingly generalizable as non-overlapping aspects are removed. If the degree of similarity is insufficient for surpassing the threshold for representation combination, these new exemplars are then compared against existing exemplars to create a new abstraction. If the new exemplar is too dissimilar from any item in memory, it is temporarily stored as a separate exemplar, which may or may not eventually get assimilated into an abstraction. Additionally, projection of candidate inferences from abstractions to exemplars increases the detail of exemplar representations. In this way, SAGE begins with no generalizations, but bootstraps its way towards abstraction via comparing exemplars and performing category-based induction.

Thus, structure mapping not only describes analogical reasoning, but it also provides a general model for cognitive development through analogical inference. Convergent support for this model is provided by its strong resemblance to the "memory-prediction framework" from computational neuroscience (Hawkins & Blakeslee, 2004), as well as the "hierarchical Dirichlet process" from machine learning (Griffiths, Canini, Sanborn, & Navarro, 2009). Both Bayesian and analogical inference allow learners to discover category structures from limited observations, and enable increasingly sophisticated representations to develop with experience. However, an unconstrained search of solution space would require examining a prohibitively large number of permutations before "toeholds" could be discovered for these bootstrapping processes. As we will discuss in the next sections, a cybernetics perspective clearly implicates embodiment as a necessary source of useful inductive biases for self-organizing cognitive systems.

## Principles of cybernetics

*"SOCRATES: Or again, in a ship, if a man having the power to do what he likes, has no intelligence or skill in navigation [αρετης κυβερνητικης, aretes kybernetikes], do you see what will happen to him and to his fellow-sailors?"*

> *–Plato, Alcibiades I*

*Cybernetics* is the interdisciplinary science of abstract principles in the organization and functioning of complex systems, emphasizing how information, models, and control actions can be used to steer towards and maintain goals, while counteracting disturbances (Heylighen & Joslyn, 2003). This analysis approach is concerned with system properties that are independent of specific components or material constituents. In this way, superficially dissimilar systems can be described with similar concepts, thus enabling comparison, synthesis, and inference generation based on a variety of source phenomena. Thus, cybernetics focuses on relations such as differences, connections between elements or processes, and transformations such that generalizable knowledge can be abstracted from physical aspects of systems.

In cybernetic analysis, the observer begins by conceptually separating the object of study (i.e., the system) from other phenomena (i.e., the environment), and proceeds to distinguish between the presence and absence of various properties of the system, which can consist of physical attributes or distinctive features of processes. The conjunction of all values of properties at a particular moment constitutes a system's state. A system is constrained to the degree that the actual variety of states that the system can exhibit is smaller than the variety of possible states. Constraints reduce uncertainty about the state of systems, and thus allow for non-trivial predictions about aspects of systems based on formally modeled relations, dependencies, or couplings. By adding more constraining information about the state of a system via observations and candidate model structures, the probabilities of alternative states become smaller, and thus model-variety and uncertainty can be reduced.

In cybernetic terms, nervous systems evolved as control systems for guiding the behavior of living organisms, which are considered as autonomous agents that pursue goals while resisting obstructions that would cause deviations from goal attainment. Since organisms are dynamical systems, these goals are equilibria in state space to which a system returns after perturbations. These goals can also be subsets of acceptable states (i.e., attractors), with the defining dimensions being essential variables that must be kept within a limited range compatible with the survival of the system. The essential variables that allow an organism to preserve itself define its homeostatic boundaries, which both constrain and are maintained by dynamic equilibria that constitute the processes of life. The equilibria are dynamic because multiple constraints need to be satisfied for an organism to survive— and many of these factors depend on conditions over which the organism only has limited control—hence it will need to shift between and balance various goals, depending on circumstance.

Varela (1979), described organisms as *self-producing* (i.e., *autopoietic*) systems (Rudrauf, Lutz, Cosmelli, Lachaux, & Le Van Quyen, 2003):

> "An autopoietic system is organized (defined as a unity) as a network of processes of production (transformation and destruction) of components that produces the components that: 1) through their interactions and transformations continuously regenerate and realize the network of processes (relations) that produce them; and 2) constitute it (the machine) as a concrete unity in the space in which they exist by specifying the topological domain of its realization as such a network."

Although functional relations are asymmetric in a variety of ways, the organism as a whole consists of mutually embedded subsystems that *couple* with each other—i.e., reciprocal action defined by simultaneous acting on and acting upon—so that the organism can more effectively couple with its environment in ways that allow sustained homeostasis, such that the process of *autopoesis* can continue. Thus, nervous systems are inherently *embodied*, and interactions with the *organizationally enclosing* bodies as well as external environments are essential for understanding their functioning.

The selective pressures producing nervous systems and other biological adaptations can be broadly classified into two categories: 1) maintaining homeostatic boundaries, and 2) ensuring reproduction. Organisms need to survive long enough to reproduce, but genetic perpetuation is the core of natural selection (Dawkins, 2006). Different genetic combinations result in different *emergent*—i.e., the *generation* of *qualitatively different wholes* from *interacting parts*—behaviors, and the process of natural selection will favor whatever maximizes overall replication of the underlying genes. These *goals* can be viewed as gradients, or "*fitness*" functions over *state spaces*, which define the degree of "*value*" or "*preference*" for different states. Thus, evolutionary processes influence nervous systems to facilitate action-patterns that maximize "*utility functions*" for genetic fitness (Dawkins, 1996).

Adapted nervous systems act as *control systems* that resist perturbations to essential variables, whether or not these perturbations originate from inside or outside of organisms (Heylighen & Joslyn, 2003). In this way, the motivations of individual organisms correspond to the "goals" of genetic evolution. There are three primary mechanisms by which these goals can be achieved through adaptive regulation: 1) *Buffering* passively absorbs or dampens perturbations; 2) *Feedforward control* suppresses disturbance before they can affect the system's essential variables; and 3) *Feedback control* compensates for *errors* or deviations from goals after they have occurred. When considered independently, none of these methods are sufficient for successful control of a complex system in a complex environment:

1. *Buffering* is achieved both by the overall robustness of the system, as well as the default reactivity of the organism to perturbations (e.g. homeostatic set points), but on its own it is incapable of handling major perturbations or returning the organism to equilibrium.

2. *Feedforward control* is achieved by actions based on the prediction of future perturbations (i.e., allostasis), but on its own it will be incapable of handling all of the permutations of disturbances that the organism may encounter.
3. *Feedback control* is achieved by responding to deviations from desired states, but it must first allow errors to occur before corrective action can be taken.

Complex nervous systems, however, incorporate all three methods on multiple levels in permitting organisms to more effectively navigate their environments, respond to challenges, and prevent entropy from internally increasing. By remembering the history of feedback control in patterns of neural connections, and then using those memories to change buffering set points and guide feedforward control, nervous systems become sufficiently effective at regulation that they are able to justify their substantial metabolic costs (Attwell & Laughlin, 2001; Howarth, Gleeson, & Attwell, 2012). According to Ross Ashby's "*law of requisite variability*," the greater the variety of disturbances that a system is likely to face, the greater the variety of actions needed by the regulator to respond to those challenges. This principle is the major factor underlying the accumulation of nervous system complexity over evolutionary time. Indeed, nervous systems are ideally suited for meeting this need, since they are able to use experience to reconfigure their connections and produce a wide range of actions.

Even extremely simple nervous systems can be thought of as predictive control systems. The flexible connectivity of neurons allows organisms to achieve all three forms of regulatory control to some extent. For example, associative learning has been demonstrated in the nematode, *Caenorhabditis elegans*, even though it only has 932 genetically-determined cells, of which 302 are neurons (Catharine H, 2004). However, association only allows an organism to achieve a limited degree of feedback (via classical conditioning) and feedforward (via stimulus generalization) control.

Nonetheless, even with a limited repertoire of actions, association is an implicit instantiation of Bayesian-analogical inference (Courville, Daw, & Touretzky, 2006).[6] More specifically, a stimulus-response pairing may have been adaptive in the past, but applying a similar response to an associated stimulus will have varying likelihoods of being adaptive outside of the original learning context. The estimated utility of a response depends on combining conditional probabilities for the degree of association between stimuli, as well as the likelihood that the responses will be appropriate. In addition to utilizing probabilistic inference, associative learning can be thought of as a basic type of analogy, in that representations corresponding to past stimulus-response feature conjunctions are compared with representations generated by present and anticipated experiences. By applying similar responses to similar stimuli, organisms use implicit analogical inference to increase the probability of adaptive behavior.

In addition to basic association, neural systems have evolved to implement *control hierarchies* (Fuster, 2009; Krawczyk, Michelle McClelland, & Donovan, 2011), where *higher-level goals* control the settings for *subsidiary goals* (Heylighen & Joslyn, 2003). By adding an additional layer of control, a goal can become the results of an action taken to achieve a higher-level goal, and more layers can be introduced for further nesting of goals and actions. Additional layers can reduce the variety of perturbations, but will not be able to eliminate errors. The number of levels required to reduce error to "acceptable" levels—which would be determined by the ability of events to impact essential variables—depends on the efficacy of each level in predicting disturbances and generating appropriate response.

---

[6] Even individual synapses implement conditional probabilities by influencing the degree of inter-neuron functional connectivity.

However, these new levels come at the expense of additional regulatory machinery, and may introduce noise or delays for some actions. In both engineering and evolution, parsimony of construction is superior, all else being equal. But when all else is not equal, the introduction of an additional control level represents a *"metastable transition,"* which is thought to be the basic unit of the evolution of cybernetic systems. In the next section, we will discuss a particularly significant metastable transition represented by the evolution of mammalian cortex.

## Cortical control systems

*"If the inputs to a system cause the same pattern of activity to occur repeatedly, the set of active elements constituting that pattern will become increasingly strongly interassociated. That is, each element will tend to turn on every other element and (with negative weights) to turn off the elements that do not form part of the pattern. To put it another way, the pattern as a whole will become 'auto-associated'. We may call a learned (auto-associated) pattern an engram."*

> *–Gordon Allport (1987-1967)*

Cortex allows organisms to remember specific sequences of events from the past, and on the basis of these memories, predict specific sequences that might occur in the future. In Fuster's (2006) model of cortex, memory and knowledge are represented in distributed and overlapping networks of interacting neurons that are "heterarchically" organized into functional cycles within perception-action hierarchies (Fuster, 2006, 2009; Fuster & Bressler, 2012). He refers to these auto-associated representational units as "cognits," and considers them to be the elemental basis of cortical memory. Over the course of development, increasingly sophisticated representations evolve as more complex patterns are abstracted from simpler patterns, which allow new kinds of patterns to be discovered. From this perspective, nested dynamics of causal relations in the world are mirrored by nested dynamics of neuronal representations, which exist in varying degrees of abstraction and complexity with respect to perception-action sequences.

Intuitively, hierarchical organization suggests that this gradient of increasing abstraction is reflected by locations of representations relative to primary sensory and motor cortices. That is, concrete perception-action sequences will be located closer to where sensations first input, and where specific motor commands first output. More abstract patterns (e.g. high-level goals) will be located closer to multi-modal association cortices and further from primary cortices. However, these abstract patterns will be connected to representations on multiple levels of the hierarchy on the basis of auto-associative links formed through past and ongoing experience.

Converging support for this theory is beginning to accumulate from the fields of computational neuroscience and machine learning. More specifically, it has been proposed that cortex may implement a common algorithm for hierarchical pattern abstraction, which affords complex associations that would be difficult to discover by statistical learning within non-hierarchically organized networks (Friston, 2008; Friston & Kiebel, 2009; George & Hawkins, 2009; Hawkins & Blakeslee, 2004; Kiebel, Daunizeau, & Friston, 2008; Schofield et al., 2009). In addition to its theoretical parsimony, this hypothesis is supported by studies of cortical micro-circuitry. After discovering the existence of cortical columns in 1978, Mountcastle suggested that these structures possibly enabled a common computational process to take place throughout cortex (Mountcastle, 1978). More recently, machine learning systems are being developed in which hierarchically organized cortical columns constitute nodes within probabilistic networks for advanced pattern recognition (Hawkins, 2011).

Cortex is anatomically distinguished by groups of approximately 80-100 neurons with common developmental origin, arranged into columns that are positioned perpendicularly to the cortical sheet with its six semi-distinct layers (Krieger, Kuner, & Sakmann, 2007; Mountcastle, 1997). Except for a few differences in visual and motor cortices, these "minicolumns" show remarkable anatomical homogeneity across brain areas, as well as across different mammalian species. Notably, cytoarchitectonic studies suggest that morphological variation may be initially lacking, and diversity is created by experience (Buxhoeveden & Casanova, 2002; E. G. Jones & Rakic, 2010). Minicolumns further self-organize into groups of approximately 60-100 units to form macrocolumns (Markram, 2006). These macrocolumns, or "cortical modules" consist of multiple minicolumns bound together via short-range connections with similar, although still heterogeneous, response properties.[7] Although the total number of cortical modules for an average human brain is difficult to determine precisely, estimates range from one to two million macrocolumns (Johansson & Lansner, 2007).

Importantly, the functional significance of a particular cortical column or neuron emerges as a function of the particular experiences of an individual organism. In light of the central role of self-organization in shaping cortex, it is unlikely that natural selection was able to influence organisms through genetically specified cortical representations, or responses to those representations. A priori, the non-linearity of development suggests that genetic selection would be unable to produce such adaptations, even with strong selective pressures. If a mechanism relies on pre-specifying details of biology that are unpredictable in principle (Elbert et al., 1994; Korn & Faure, 2003; Rabinovich & Abarbanel, 1998; Wolfram, 2002), then it is also un-evolvable. Thus, rather than "hard-wiring" particular adaptive behaviors, cortex evolved as a general-purpose learning system that attempts to predict future observations on the basis of past experiences.

---

[7] Column-based cortical modules have no relation or resemblance to the special-purpose computational modules discussed within evolutionary psychology.

From this perspective, Bayesian-analogical inference is an emergent property of a common cortical algorithm that shapes sequences of representations so that organisms can more effectively couple with their environments. In terms of Bayesian probability, goal oriented behavior would result from a particular kind of semi-parallel, semi-stochastic, self-sampling generative model, where nodes corresponding to specific action dynamics are selected based on model-estimated utility. In terms of analogical structure mapping, this biased search of perception-action sequence representations would attempt to maximize alignment between estimated—current states as well as the consequences of potential actions—and desired experiences by chaining together multiple representations in varying configurations, at multiple levels of abstraction. Since different sequence representations correspond to different imagined and physically generated behaviors, goal attainment could be achieved by maximizing the overall degree of alignment between the representations activated by imaging desired states and representations activated by imagining continuing and potential actions.

More specifically, if cortex automatically performs structure mapping among the most highly activated networks, then analogical alignment of representations could adaptively select from a hierarchy of action sequences at different levels of abstraction. That is, by imagining different scenarios relative to goal attainment, alignable differences would correspond to subordinate actions, for which alignable differences would represent further subordinate actions within a hierarchy of action sequences. As each potential action is considered, the auto-associative property of nervous systems would automatically activate candidate action representations based on similarities with past experiences. Considering that superordinate and subordinate action sequences are themselves auto-associatively linked, they would provide mutual constraints as parallel structure mapping continuously minimizes overall alignable differences between representations on multiple levels. By this account, similar cognitive processes could be involved in selecting high-level strategies for goal attainment, as well as in low-level adjustment of specific sequences retrieved from memory in a moment-to-moment fashion for precise motor control.

Theoretically, cortex could efficiently generate utility maximizing analogical inferences by minimizing through a general principle of "free energy" minimization (Friston, 2010; Hawkins, 2011; Kozma, Puljic, Balister, Bollobas, & Freeman, 2004). Multiple compelling mechanisms have been proposed within this paradigm, and here we will propose a novel mechanism by which the brain could tune itself to minimize its overall prediction-error and implement credit-assignment during learning:

1. If a minicolumn's inputs are predicted in advance via stimulation of specific inhibitory interneurons within the column, then only those neurons without their respective inhibitory interneurons activated will increase their firing rates.
2. However, if a sufficient number of non-predicted inputs occur, and a percolation threshold is surpassed, the entire column will become active, resulting in a cascade of activity-predictions in functionally connected columns.
3. Depending on the degree of functional connectivity with inhibitory interneurons of neuromodulatory nuclei—where changes in the activity of a single neuromodulator releasing neuron can influence millions of cortical neurons through their diffuse

projections—any dynamic that causes overall activity to be reduced should result in decreased inhibition of the production of these neuromodulators.

4. This net disinhibition would enhance the most robustly active patterns, strengthen the connections underlying these patterns (i.e., reinforcement), and thus increase the efficiency of the dynamics contributing to successful prediction (i.e., minimized error signals).

Although this activity-minimizing algorithm could potentially result in stasis, regulatory nuclei of the hypothalamus and brainstem will stimulate these inhibitory interneurons to the degree that action is needed to restore homeostatic balances. Thus an organism could not remain permanently inactive, as physiological signals of hunger would result in stimulation of these regulatory nuclei, whose activity can be thought of as signifying the distance from homeostatic set points, or as signifiers of biologically specified predictions for which deviations result in error signals. Over time, cortical dynamics resulting in the minimization of error signals from these regulatory nuclei will become distributed across the cortical heterarchy as habitual predictions. The impact of these habitual predictions on overall functioning would constitute the evolving utility function of the organism.

In this way, although natural selection is limited in its ability to influence specific behaviors, utility could be maximized—within bounds—by the tendency for dynamical systems to minimize free energy. In terms of analogy, this would correspond to minimizing the complexity of the underlying sequence representations. However, the body provides an initial set of values that constrains which of the countless aspects of the generative model are optimized. Indeed, BAC suggests that organisms are such effective learners because they begin with a sense of their own embodiment as a kind of prototypical object from which they can partially generalize, and they pay attention to this object because it is directly connected to the mechanisms of reinforcement (Lekmes & Tracey, 2008). This account of cognitive development can potentially explain the knowledge that infants possess (Carey & Gelman, 1991), as well as how they use it to partially generalize to other classes of phenomena (Lakoff & Johnson, 1999), even if sometimes incorrectly.

Thus, Bayesian-analogical inference not only allows organisms to interpret ambiguous information from impoverished sensory data, but it also enables modulation of behavior based on prior knowledge. To the degree that previous experiences allow organisms to imagine outcomes as well as anticipated reward value, they are able to make choices that exhibit bounded rationality relative to current preferences and beliefs, which can then be updated through further experience. In this way, cortical memory automatically shapes behavior on the basis of evolving expected utility estimates. Additionally, as reasoning becomes increasingly sophisticated, high-level cognition can contribute to the types of expectations that learners will form, which in turn determine what they will do, experience, and come to expect in the future. As the developing person becomes capable of perceiving reward-value from increasingly abstract reinforcers, these high-level meanings would be crucial for understanding the future course of cognitive-affective evolution.

## BAC and consciousness

*"The aim of scientific explanation throughout the ages has been unification, i.e., the comprehending of a maximum of facts and regularities in terms of a minimum of theoretical concepts and assumptions."*

> *–Herbert Feigl (1902-1988)*

As a partial test of the BAC framework, we will now attempt to parsimoniously describe aspects of a poorly understood phenomenon: how neural systems give rise to conscious experiences. Although embodiment is essential for enabling self-awareness and higher-level properties of consciousness (A. D. Craig, 2003; A. D. B. Craig, 2009; Damasio, 2000, 2003), in this particular discussion we will focus on the role of Bayesian-analogical inference in supporting the capacity for conscious awareness.

According to the "integrated information" theory of consciousness, qualitative properties of experience (i.e., qualia) are generated by relationships of activity within neuronal complexes (Balduzzi & Tononi, 2009; G Tononi & Edelman, 1998). These "quale states" correspond to a dynamically changing core of functional activity, which can be analyzed as topological manifolds within abstract feature-spaces. This promising theory may help to explain many properties of the brain, such as hierarchical organization, small-world network connectivity, and localization of function. Indeed, information-theoretical measures of effective computational capacity may be able to characterize network properties that are necessary for supporting various kinds of cognitive processing in different systems.

In systems where complexes produce a sufficient degree of integrated information, numerous and varied processes can be synergistically unified in a "dynamic core" of neuronal ensembles (Gerald M Edelman, Gally, & Baars, 2011; Seth & Baars, 2005). The dynamic core is a complex defined by coalitions of patterns that "compete" and "cooperate" within the brain (G. M. Edelman, 1987, 1993). Patterns contributing to core dynamics can undergo re-entrant feedback amplification from the mutually constraining relations of the patterns constituting the grand coalition, which are stabilized via their functional interconnectivity. This emergent 'virtual machine' creates a system of relations, which temporarily increases the amount of integrated information available to patterns that can successfully couple with the dynamic core. In this way, complexes with a critical mass of integrated functional connectivity enable a "global workspace" that allows information to be more effectively shared among distributed neural systems (Baars, 2005). Thus BAC proposes that consciousness is the capacity of the mind to support global workspaces defined by a dynamic core of competing patterns, which depends on, but is not identical to a system's integrated information.

By equating consciousness with integrated information, Tononi et al. (2008) may be conflating necessary with sufficient conditions. Further, the emphasis on mechanistic realization may overlook "real patterns" on the cognitive level (D. Dennett, 1991, 1996), for which the relevant details are more effectively described in terms of Bayesian-analogical inference. For a system to be 'conscious', integrated information must apply to representations with experience-grounded meanings. These representations may not necessarily be explicitly defined symbols, but their semiotic content could be implied in a cybernetic manner, based on the ways in which particular features of the system influence how it interacts with other systems. A complex could have an arbitrarily high amount of integrated information, but it will not be conscious unless it also refers to patterns external to the system. More accurately, a system's capacity for consciousness is proportional to the integrated information that can impact representational dynamics.

With respect to dynamic core theory, Edelman (2004) may have been mistaken in stipulating that conscious states are not causal, but constitute *epiphenomenal* consequences of core transformations (G. Edelman, 2004). First, 'causality' is not just a simple logical primitive, or some 'objective' feature of nature. Rather, it is an abstraction, or an analogy with a moderate degree of metaphoricity relative to observations whose representations achieve a sufficient degree of similarity that we categorize them as "causal." From a BAC perspective, causality is a high-level schema for processes that tend to change in regular and predictable ways according to factors such as temporal ordering and spatial continuity. Further, complex phenomena are best understood in terms of *emergent causality*, wherein system-wide properties are both influenced by the synergistic interactions of its constituents, but these interactions are also influenced by system wide properties. Or, in other words, neither parts nor wholes are ontologically primary in any absolute sense; rather, their functional interdependence means that wholes and parts are inseparable. Indeed, it may be possible to formalize this sort of reciprocal causation using the same self-recursive functions that characterize all dynamical non-linear (i.e., chaotic) systems, including minds. Thus, consciousness exhibits emergent causality in that is capable of constraining and being constrained by cognitive processes impacted by the dynamic core.

In light of these considerations, not only are probabilistic analogies fundamental to all cognition, but the inferential power of explicit analogical reasoning may have been an important selective pressure for the adaptations underlying human-like consciousness (Gentner, 2010). This hypothesis is supported by analyses of long distance brain connectivity where prefrontal cortex was found to be a topologically central "hub" with a disproportionately large number of vertices in a network model (Modha & Singh, 2010). Although these data were based on tracing studies in rhesus monkeys, this prefrontal hub has been maximally increased in humans relative to non-human primates (Preuss, 2011). Further, prefrontal activation is strongly implicated in analogical reasoning, which is uniquely developed in humans, and which almost certainly depends on coordinating dynamics in multiple brain areas (Bunge, Wendelken, Badre, & Wagner, 2005; Cho et al., 2010; Green, Fugelsang, Kraemer, Shamosh, & Dunbar, 2006; Krawczyk, McClelland, Donovan, Tillman, & Maguire, 2010; Luo et al., 2003; R. G. Morrison et al., 2004; Preusse, van der Meer Elke, Deshpande, Krueger, & Wartenburger, 2011; Volle, Gilbert, Benoit, & Burgess, 2010).

Additionally, through reentrant stabilization of representations participating in the dynamic core, consciousness may be a necessary adaptation for the kind of analogical reasoning that seems to have been uniquely developed in humans (Gentner, 2010; Premack, 1983). By privileging relational information and high-level conceptual structures (e.g. causal models), learners are capable of inferring analogies that would be undiscoverable on the basis of mere perceptual similarity. Furthermore, structured representations can also increase the integrated information of the core by minimizing failed-predictions, thereby helping to stabilize complex dynamics within the grand canonical ensemble of the mind. Thus consciousness and analogy are both mutually supporting and mutually constraining in their synergistic inter-relationships.

Finally, this reciprocal causation implies that abstract cognition can alter core dynamics on the basis of moment-by-moment transformations of explicit meanings and simulated experiences. Since semantic properties can specify functional connections between disparate representations within the cortical hetearchy, these cognitive processes—and the neural adaptations supporting large-scale information integration—vastly increase the number of permutations for possible functional interconnections. In this way, conscious cognition may flexibly constrain cortical generative models as they estimate the latent variables of reality.